\documentclass{aastex62}
\begin{document}
\title{KIC 10736223: An Algol-type eclipsing binary just undergone the rapid mass-transfer stage}
\correspondingauthor{Xinghao Chen, XiaoBin Zhang, Yan Li}
\email{chenxinghao@ynao.ac.cn; xzhang@bao.ac.cn; ly@ynao.ac.cn}
\author{Xinghao Chen}
\affiliation{Yunnan Observatories, Chinese Academy of Sciences, P.O. Box 110, Kunming 650216, China}
\affiliation{Key Laboratory for Structure and Evolution of Celestial Objects, Chinese Academy of Sciences, P.O. Box 110, Kunming 650216, China}
\author{Xiaobin Zhang}
\affiliation{Key Laboratory of Optical Astronomy, National Astronomical Observatories, Chinese Academy of Sciences, Beijing, 100012, China}
\author{Yan Li}
\affiliation{Yunnan Observatories, Chinese Academy of Sciences, P.O. Box 110, Kunming 650216, China}
\affiliation{Key Laboratory for Structure and Evolution of Celestial Objects, Chinese Academy of Sciences, P.O. Box 110, Kunming 650216, China}
\affiliation{University of Chinese Academy of Sciences, Beijing 100049, China}
\affiliation{Center for Astronomical Mega-Science, Chinese Academy of Sciences, 20A Datun Road, Chaoyang District, Beijing, 100012, China}
\author{Hailiang Chen}
\affiliation{Yunnan Observatories, Chinese Academy of Sciences, P.O. Box 110, Kunming 650216, China}
\affiliation{Key Laboratory for Structure and Evolution of Celestial Objects, Chinese Academy of Sciences, P.O. Box 110, Kunming 650216, China}
\author{Changqing Luo}
\affiliation{Key Laboratory of Optical Astronomy, National Astronomical Observatories, Chinese Academy of Sciences, Beijing, 100012, China}
\author{Jie Su}
\affiliation{Yunnan Observatories, Chinese Academy of Sciences, P.O. Box 110, Kunming 650216, China}
\affiliation{Key Laboratory for Structure and Evolution of Celestial Objects, Chinese Academy of Sciences, P.O. Box 110, Kunming 650216, China}
\author{Xuefei Chen}
\affiliation{Yunnan Observatories, Chinese Academy of Sciences, P.O. Box 110, Kunming 650216, China}
\affiliation{Key Laboratory for Structure and Evolution of Celestial Objects, Chinese Academy of Sciences, P.O. Box 110, Kunming 650216, China}
\author{Zhanwen Han}
\affiliation{Yunnan Observatories, Chinese Academy of Sciences, P.O. Box 110, Kunming 650216, China}
\affiliation{Key Laboratory for Structure and Evolution of Celestial Objects, Chinese Academy of Sciences, P.O. Box 110, Kunming 650216, China}
\affiliation{University of Chinese Academy of Sciences, Beijing 100049, China}
\affiliation{Center for Astronomical Mega-Science, Chinese Academy of Sciences, 20A Datun Road, Chaoyang District, Beijing, 100012, China}



\begin{abstract}
This paper reports the discovery of an Algol system KIC 10736223 that just past the rapid mass transfer stage. From the light curve and radial-velocity modelling we find KIC 10736223 to be a detached Algol system with the less-massive secondary nearly filling its Roche lobe. Based on the short-cadence Kepler data, we analyzed intrinsic oscillations of the pulsator and identified six secured independent $\delta$ Scuti-type pulsation modes ($f_{1}$, $f_3$, $f_{9}$, $f_{19}$, $f_{42}$, and $f_{48}$). We compute two grids of theoretical models to reproduce the $\delta$ Scuti freqiencies, and find fitting results of mass-accreting models meet well with those of single-star evolutionary models. The fundamental parameters of the primary star yielded with asteroseismology are $M$ = $1.57^{+0.05}_{-0.09}$ $M_{\odot}$, $Z$ = 0.009 $\pm$ 0.001, $R$ = $1.484^{+0.016}_{-0.028}$ $R_{\odot}$, $\log g$ = $4.291^{+0.004}_{-0.009}$, $T_{\rm eff}$ = $7748^{+230}_{-378}$ K, $L$ = $7.136^{+1.014}_{-1.519}$ $L_{\odot}$. The asteroseismic parameters match well with the dynamical parameters derived from the binary model. Moreover, our asteroseismic results show that the pulsator is an almost unevolved star with an age between 9.46-11.65 Myr for single-star evolutionary models and 2.67-3.14 Myr for mass-accreting models. Thereofore, KIC 10736223 may be an Algol system that has just undergone the rapid mass-transfer process.
\end{abstract}
\keywords{Asteroseismology - binaries: eclipsing - stars: individual (KIC 10736223) - stars: oscillations - stars: variables: $\delta$ Scuti}
\section{Introduction}
The pulsating stars in eclipsing binaries are particular attractive objects since the synergy can allow us directly determine accurate stellar fundamental parameters and offer us significant insight into the interiors of the star(Aerts \& Harmanec 2004; Mkrtichian et al. 2005, 2007). There are a handful of efforts to investigate interior physics of stars in details by simultaneous binary and asteroseismic modelling, such as for solar-like pulsating binaries (White et al. 2017; Beck et al. 2018; Li et al. 2018; Themassl et al. 2018;) and g-mode pulsating binaries (Hambleton et al. 2013; Keen et al. 2015; Schmid \& Aerts 2016; Guo et al. 2016; Johnston et al. 2019). Beck et al. (2018) presented an asteroseismic study of the eccentric binary system KIC 9163796 that includes two oscillating red-giant stars, and found the two component stars being in the early and late phase of the first dredge-up event on the red giant branch. Schmid \& Aerts (2016) combined binary and asteroseismic modelling for the $\delta$ Scuti-$\gamma$ Doradus hybrid binary KIC 10080943, and found that the amount of core overshooting and diffusive mixing can be well constrained under the assumption of equal age for the two stars. Besides, pulsating stars in post-mass transfer binary also deserve more attention, which carry evidences of binary interaction, offering us the opportunity to refine the theory of stellar structure and evlution. Guo et al. (2017) analyzed a post-mass transfer hybrid pulsator in eclipsing binary KIC 9592855 and found that rotation rates of both the core and envelope are similar to that of the orbital motion. Bowman et al. (2019) analyzed the TESS light curve of the oEA system U Gru and discovered that tidally preturbed oscillations also occur in p-mode region. At present, about two hundred of eclipsing binaries have been found to contain $\delta$ Scuti-type pulsating components (Kahraman Alicavus et al. 2017; Liakos $\&$ Niarchos 2017; Gaulme \& Guzik 2019). However, due to the low radial orders, mode identifications are very difficult for $\delta$ Scuti pulsators, the studies of those systems mostly concentrate on binary properties (da Silva et al. 2014; Zhang et al. 2015; Wang et al. 2018; Lee et al. 2019), comprehensively asteroseismic modelling of $\delta$ Scuti pulsators in eclipsing binaries are still lack.

KIC 10736223 is an Algol-type eclipsing binary with an orbital period of 1.1051 days, discovered by Dahlmark (2000) in Cygnus. In the Kepler Input Catalogue, the effective temperature $T_{\rm eff}$ and the metallicity of [Fe/H] for the star were given to be 7797 K and -0.193 (Pinsonneault et al. 2012), respectively. The low-resolution spectroscopy observed by the LAMOST project (Luo et al. 2012) while yields an effective temperature of 7554 K and the metallicity of [Fe/H] = -0.511 for the star. Besides, Matson et al. (2017) obtained a set of radial velocity measurements for KIC 10736223, and determined masses of the two component stars to be 1.6 $\pm$ 0.1 $M_{\odot}$ and 0.35 $\pm$ 0.03 $M_{\odot}$, respectively. Moreover, Debosscher et al. (2011) surveyed the variability of KIC 10736223 and found three main frequencies at 1.810503 day$^{-1}$, 0.905251 day$^{-1}$, and 4.523270 day$^{-1}$. These frequencies are 2, 1, and 5 times of the orbital frequencies, respectively. The intrinsic pulsations of KIC 10736223 are still poorly studied.

In this work, we carry out a comprehensive analysis for the eclipsing binary system KIC 10736223, including eclipse analysis and asteroseismic modelling. In Section 2, we describe how physical parameters of KIC 10736223 are determined through the analysis of light and radial-velocity curves. In Section 3, we analyse pulsation characteristics of the primary star. In Section 4, we elaborate details of input physics and our asteroseismic modeling. Finally, we summarize and discuss the results in Section 5.
\section{The Eclipsing Binary System}
\begin{figure*}[!h]
\includegraphics[width=0.75\textwidth, angle = -90]{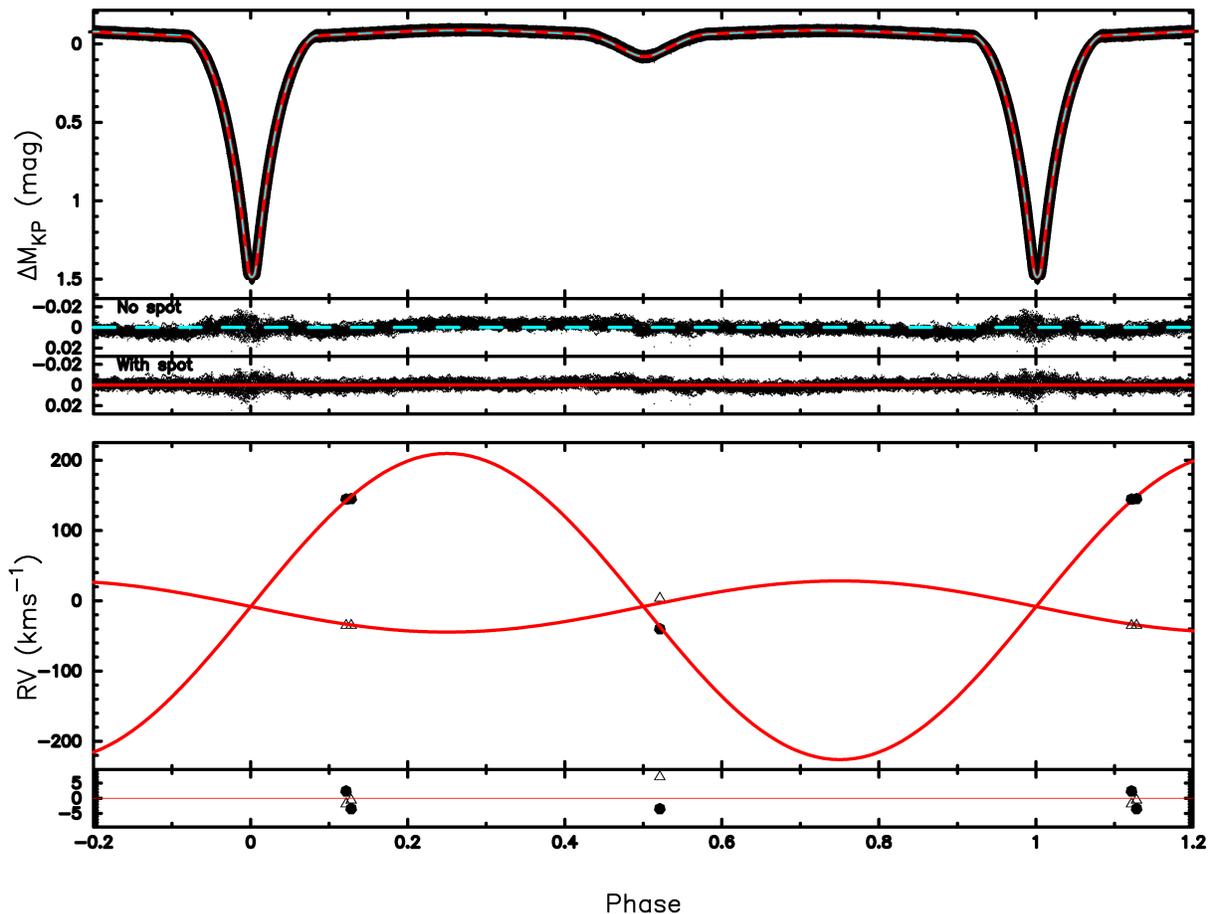}
\caption{\label{Figure 1}Light and radial velocity curves of KIC 10736223. The upper panel presents the light curve and the O-C residuals (black dots), and the lower panel presents the radial velocity curve and the O-C residuals. The solid hexagon and the triangle denote the measured radial velocities of the primary and secondary star, respectively. In the upper panel, the green dashed line denotes the unspotted synthesis curve, and the red solid line represents the spotted synthesis curve. The red lines in lower panel denote the synthesis curve of the radial velocities.}
\end{figure*}
\begin{table*}[!h]
\centering
\large
\caption{\label{t1}Physical parameters of the binary KIC 10736223. $^{\rm a}$ marks assumed values.}
\begin{tabular}{lcccc}
\hline\hline
Parameter             & Without spot &         With spot\\
\hline
$P_{\rm orb}$(days)      &   1.1050943        & 1.1050943    \\
$i$(deg)                         &   89.657$\pm$0.007 & 89.411$\pm$0.006\\
$q=M_2/M_1$               &    0.211$\pm$0.001       &0.1998$\pm$0.0003  \\
$T_{1}(\rm K)$      &7554$\pm$ 200$^{\rm a}$     &  7554$\pm$ 200$^{\rm a}$ \\                  
$T_{2}(\rm K)$     & 5045$\pm$ 135 &5006$\pm$135\\
$\Omega _{1}$     &4.1529$\pm$0.0005 &4.1431$\pm$0.0005\\             
$\Omega _{2}$     &2.3167$\pm$0..0013 &2.2713$\pm$0.0011\\
$L_{1}/(L_{1}+L_{2})_{Kp}$   &0.878$\pm$0.001   &0.878$\pm$0.001   \\
r$_{1}$(pole)        &0.2533$\pm$0.0001  &0.2532$\pm$0.0001          \\
r$_{1}$(point)      &0.2575$\pm$0.0001 &0.2573$\pm$0.0001         \\
r$_{1}$(side)       &0.2558$\pm$0.0001  &0.2557$\pm$0.0001        \\
r$_{1}$(back)        &0.2570$\pm$0.0001   &0.2569$\pm$0.0001     \\
r$_{2}$(pole)         &0.2230$\pm$0.0007  &0.2230$\pm$0.0006\\
r$_{2}$(point)        &0.2720$\pm$0.0020 &0.2781$\pm$0.0021\\
r$_{2}$(side)          &0.2305$\pm$0.0008  &0.2308$\pm$0.0007\\
r$_{2}$(back)        &0.2537$\pm$0.0012 &0.2560$\pm$0.0011\\\
Spot parameters:\\
\hline
Colatitude (deg) & \nodata & 33.4(3)\\
Longitude (deg) & \nodata & 200.4(5)\\
Radius (deg) & \nodata & 43.0(2) \\
$T_{\rm spot}/T_{\rm local}$ & \nodata & 0.934(7)\\
$\sum(\rm O-C)^2$ (10$^{-4}$)  &0.8876 &0.8736\\
\hline
Absolute parameters:\\
\hline
a ($R_{\odot}$)                   &  5.68$\pm$0.11                                   &5.69$\pm$0.11\\
$\gamma$ (km s$^{-1}$)    &  -5.56$\pm$1.73                                   &-6.40$\pm$1.74\\
$M_{1}(M_{\odot})$            & 1.66$\pm$0.09          &$1.69\pm0.09$    \\                              
$M_{2}(M_{\odot})$            & 0.35$\pm$0.02            &$0.34\pm0.02$\\
$R_{1} (R_{\odot}$)            &$1.45\pm0.03$              &$1.45\pm0.03$   \\                               
$R_{2} (R_{\odot}$)           &$1.34\pm0.03$  &  $1.35\pm0.03$\\
$L_{1}(L_{\odot})$                   &6.08$\pm$0.72       &6.14$\pm$0.72           \\                       
$L_{2}(L_{\odot})$          &1.04$\pm$0.12 &1.14$\pm$0.12\\
$\log g_{1}$                    &$4.34\pm0.03$                    &$4.34\pm0.03$    \\                              
$\log g_{2}$   &$3.71\pm0.03$  &$3.71\pm0.03$\\
\hline
\end{tabular}
\end{table*}
KIC 10736223 was observed in both the long cadence mode and the short cadence mode by the Kepler satellite. For the purpose of this work, we used only the short cadence data, which were obtained during the Kepler's observing quarter 16, from March 7 to April 8, 2013 with a time span of $\Delta$T = 31.8 days. We downloaded the data from the MAST data archive center (http://archive.stsci.edu/kepler/). Following the method described in Slawson et al. (2011), the simple aperture photometry light curve extracted from the data file was detrended and normalized, the outliers were removed. The radial-velocity data of the binary system are adopted from Matson et al. (2017) containing of 6 measurements, 3 for the primary and 3 for the secondary component. Using the ephemerides given in the Kepler eclipsing binary catalog (Kirk et al. 2016), phases were computed and the  light and radial-velocity curves were folded as displayed in Figure 1.

In order to obtain reliable parameters for this eclipsing binary system, we model the light and radial-velocity curves simultaneously following the way of Zhang et al. (2018) by using the Wilson-Devinney method (Wilson $\&$ Devinney 1971; Wilson 1979, hereafter W-D). The 2013 version of the W-D code was employed for light-curve analysis. In doing that, the effective temperature of the primary star was fixed at $T_{\rm eff,1}$ = 7554 K based on the LAMOST spectroscopy. The gravity-darkening exponent and the bolometric albedo of the primary star were both set to be 1.0 according to Lucy (1967) and Rucinski (1969), considering that it could have a radiative envelope. Those parameters of the secondary component were taken as 0.32 and 0.5, respectively, according to its effective temperature resulted from the iteration. The initial bolometric limb-darkening coefficients in logarithmic form ($X_{1,2}$, $Y_{1,2}$) were taken from van Hamme (1993); the monochromatic ones ($x_{1,2}$, $y_{1,2}$) in the Kepler band from Claret $\&$ Bloemen (2011) were adopted. Since the orbit of the binary is tight, the light curve presents to be symmetric with two eclipses being separated exactly by 0.5 and having equal eclipsing durations in phase, KIC10736223 ought to be a circularized and synchronized system. A synchronous rotation for both components and a circular orbit were therefore adopted for the binary model.

We started the differential-correction (DC) program of the W-D code from Mod 2 (with detached configuration). The free parameters are the mass ratio (q=$M_{2}/M_{1}$), the orbital inclination (i), the phase shift, the effective temperature of star 2 ($T_{2}$), the dimensionless surface potential ($\Omega$) and luminosity of two stars. These parameters were alternately adjusted till a converged solution was reached. The results are listed in the second column of Table 1. Based on the solution, the theoretical light curve was calculated as presented by the dashed curve in the top panel of Figure 1. The O-C light residuals of the solution are plotted in the second panel of the figure, wherein it can be seen that there is a structure remaining in the residuals which is coherently phased with the orbit. The unspotted binary model does not describe the observed light curve perfectly around phases 0.2 and 0.5.

The initial binary solution reveals that KIC 10736223 is very likely a detached system consisting of a K-type secondary component. The light curve discrepancy may result from the probable magnetic activity of the cool secondary. We thus tested the possibility by placed a cool spot on the secondary star to solve the light discrepancy. The spot parameters are the spot temperature given as a fraction of the surrounding photospheric temperature, the spot radius and its colatitude and longitude. The preliminary spot longitude could be found approximately from the phases of the light distortion. The other three parameters were calculated by adjusting the theoretical light curve to fit approximately the observed light curve. The spot parameters were then adjusted along with the free parameters. Finally, we obtained the best-fitting model. The best solution is given in the third column of Table 1 and is described by a solid curve in the top panel of Figure 1. The parameters errors given in Table 1 are produced by the DC code computed from the covariance matrix using the standard method. The bottom panel of Figure 1 presents the O-C residuals computed from the spotted binary model. It shows the cool spot on the secondary component could almost entirely explain the light variation. We calculate the sum of the squared residuals for the models with spot and without spot, respectively. As shown in Table 1, the $\sum(\rm O-C)^2$ value of the model with spot is smaller than that of the model without spot.

The synthesis to the radial velocity measurements based on the best solution is illustrated in lower two panels in Figure 1. The synthesis yields a semi-axis of 5.69$\pm$0.11 R$_{\odot}$ for the binary system. Combining the spectroscopic solution with the results of light curve modeling, the physical parameters including mass, radius, luminosity, and surface gravity of the two components were determined as given in Table 1. The physical parameters derived for the primary component suggests that it is an unevolved main-sequence star. The less-massive secondary is while much evolved. With a short orbital period and a small mass ratio, KIC 10736223 is typical a classical Algol system which is formed through mass exchange and mass-ratio reversal. With a filling factor of $R_2$/$R_{\rm cr2}$ = 0.97, the secondary star is almost filling its Roche lobe, probably indicating that the binary system just past the rapid mass-transfer stage.
\section{Frequency Analysis}
\begin{figure*}[!h]
\includegraphics[width=0.8\textwidth, angle = -90]{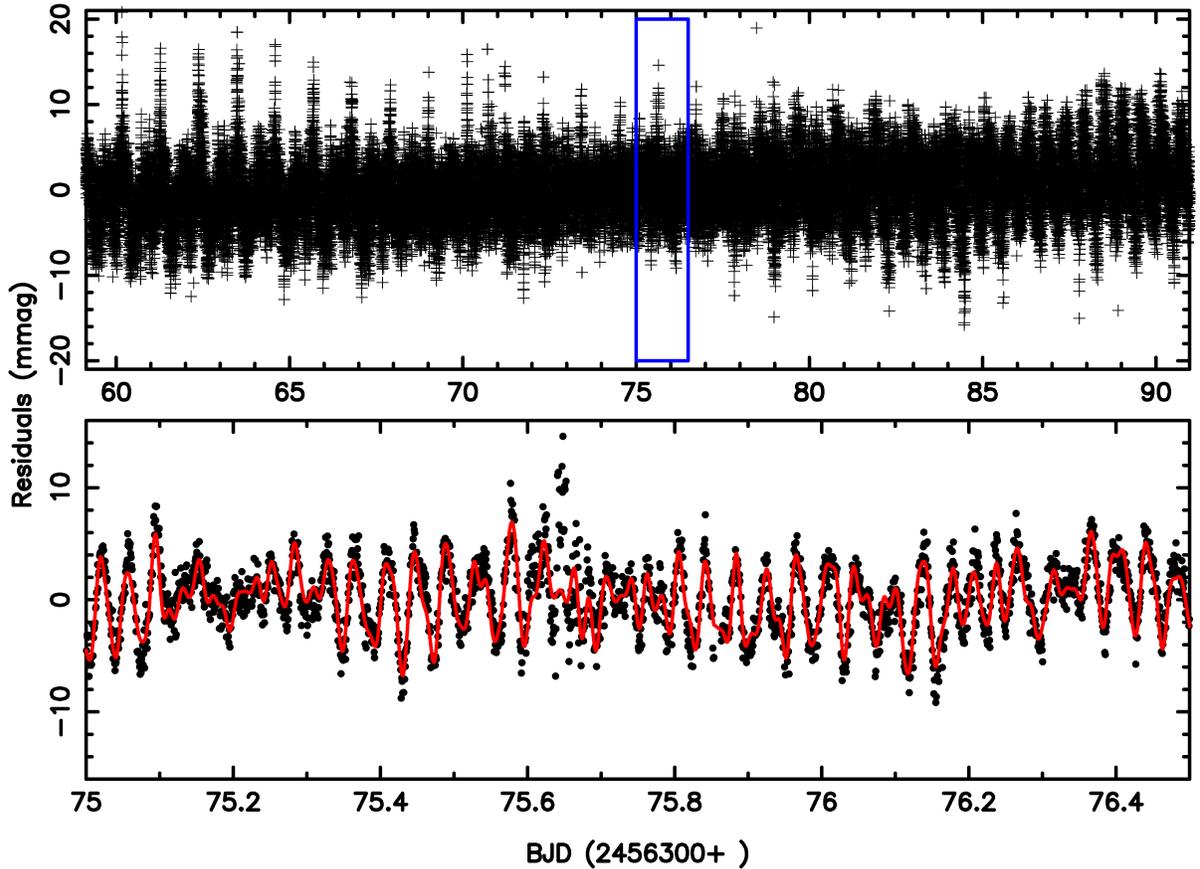}
\caption {\label{Figure2}Light  residuals after removing the eclipsing light changes from the Kepler photometric data. The lower panel presents a portion of the residuals denoted with the inset box in the upper panel. The red line indicates the synthetic curve computed using all the frequencies detected.}
\end{figure*}
\begin{figure*}[!h]
\includegraphics[width=0.8\textwidth, angle = -90]{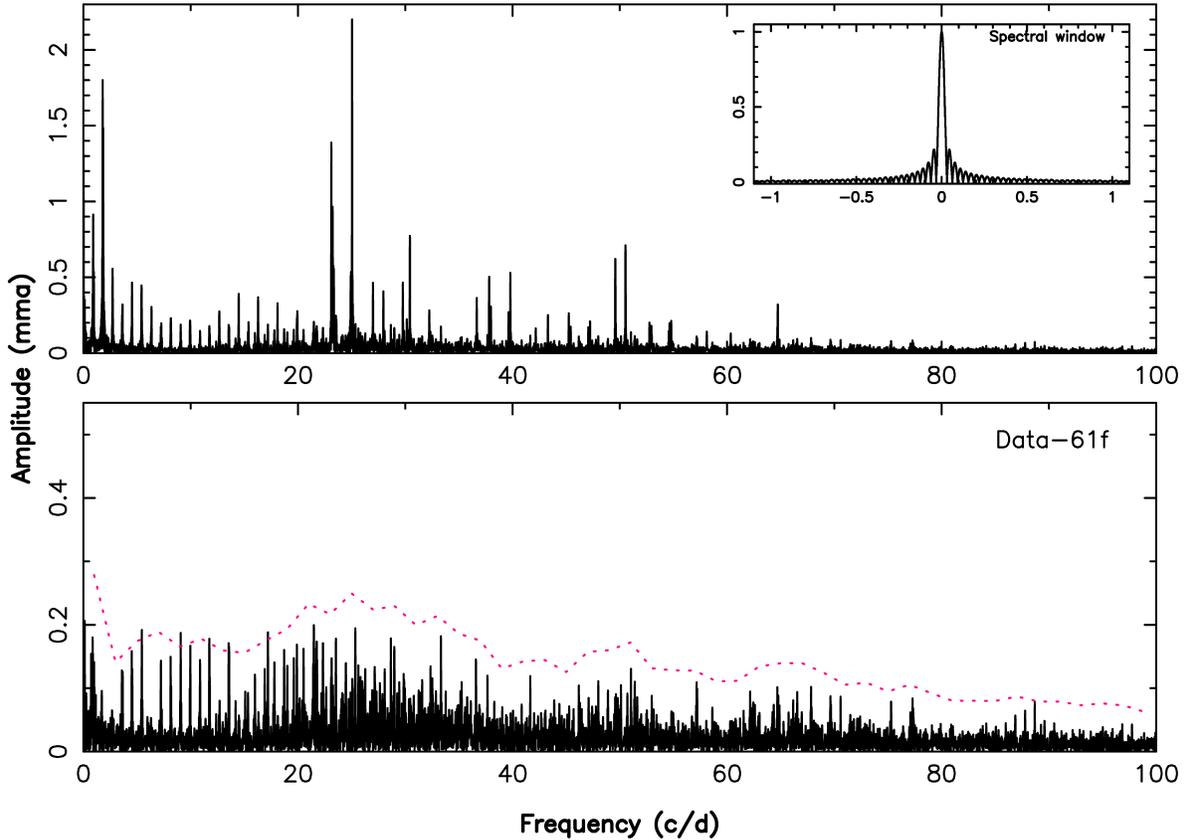}
\caption{\label{Figure 3}Fourier amplitude spectrum of the light residuals for KIC 10736223. The upper panel presents the original spectrum. The inset panel in the upper panel shows the window spectrum. The lower panel presents the residual spectrum after the 61 detected frequencies were subtracted. The dotted line in red shows the level of S/N =5.4.}
\end{figure*}
\begin{table*}[!h]
\centering
\tiny
\caption{\label{t2}Frequencies detected for KIC 10736223. The columns named ID denote the serial number of the detected frequencies. The columns named Freq. denote detected frequencies in units of cycle per day and $\mu$Hz, respectively. The columns named S/N denote their signal-to-noise. The column named Remark denote possible identifications for the detected frequencies. }
\begin{tabular}{cccccccccccc}
\hline\hline
 ID  &Freq.  &Freq.  &Ampl.  &S/N    &Remark  &ID   &Freq.   &Freq.  &Ampl.  &S/N & Remark \\
&(day$^{-1}$)   &($\mu$Hz)  &($\pm$ 11.2$\mu$mag)   & &  & &(day$^{-1}$)       &($\mu$Hz)  &($\pm$ 11.2$\mu$mag) & \\
\hline
\boldmath{$f_{ 1}$} &\bf{25.0445(1)} &\bf{289.867(1)} &\bf{2268.0} &\bf{49.0} &     &$f_{32}$ & 6.3467(6) & 73.457(7) & 303.4 &10.9 &$7f_{\rm orb}$\\
$f_{ 2}$ & 1.7999(1) & 20.832(1) &1628.7 &43.6 &$2f_{\rm orb}$                      &$f_{33}$ & 1.7732(6) & 20.523(7) & 303.7 & 8.1 &Side lobe\\
\boldmath{$f_{ 3}$} &\bf{23.1127(1)} &\bf{267.508(1)} &\bf{1370.9} &\bf{32.7} &     &$f_{34}$ &32.2456(7) &373.213(8) & 277.1 & 6.8 &$f_1+8f_{\rm orb}$\\
$f_{ 4}$ & 0.0220(1) &  0.254(2) &1084.0 &16.2 &2$f_{\rm sat}$                                    &$f_{35}$ &19.9433(7) &230.825(8) & 277.3 & 6.8 &$22f_{\rm orb}$\\
$f_{ 5}$ & 1.8392(1) & 21.286(2) &1161.1 &32.1 &Side lobe                           &$f_{36}$ &23.5839(7) &272.961(8) & 274.9 & 6.7 &$2f_3-25f_{\rm orb}$\\
$f_{ 6}$ &23.2321(2) &268.890(2) & 959.3 &22.4 &$f_1-f_2$                           &$f_{37}$ &12.6699(7) &146.642(8) & 273.4 & 9.2 &$14f_{\rm orb}$\\
$f_{ 7}$ & 0.9172(2) & 10.616(2) & 916.4 &16.9 &$f_{\rm orb}$                       &$f_{38}$ & 2.6998(7) & 31.248(9) & 261.6 & 9.0 &Side lobe\\
$f_{ 8}$ &30.4347(2) &352.254(3) & 780.7 &20.0 &$f_1+6f_{\rm orb}$                  &$f_{39}$ &39.6352(7) &458.741(9) & 261.9 &11.0 &$f_9-12f_{\rm orb}$\\
\boldmath{$f_{ 9}$} &\bf{50.5292(2)} &\bf{584.829(3)} & \bf{702.8} &\bf{20.7} &     &$f_{40}$ &45.2442(7) &523.659(9) & 259.9 &10.8 &$50f_{\rm orb}$\\
$f_{10}$ &23.3436(3) &270.180(4) & 587.0 &14.5 &$f_9-30f_{\rm orb}$                 &$f_{41}$ &43.3014(8) &501.173(9) & 249.4 & 9.7 &$f_9-8f_{\rm orb}$\\
$f_{11}$ &49.5821(3) &573.867(3) & 621.8 &20.5 &$f_9-f_{\rm orb}$                   &\boldmath{$f_{42}$} &\bf{30.1442(8)} &\bf{348.891(10)} &\bf{ 231.4} &\bf{ 6.1} &\\
$f_{12}$ &24.9597(3) &288.885(4) & 580.5 &12.6 &$f_3+f_5$                           &$f_{43}$ & 8.1529(9) & 94.362(10) & 228.1 & 8.0 &$9f_{\rm orb}$\\
$f_{13}$ & 2.7234(3) & 31.521(4) & 548.1 &19.4 &$3f_{\rm orb}$                      &$f_{44}$ & 9.9449(9) &115.103(10) & 220.4 & 7.7 &$11f_{\rm orb}$\\
$f_{14}$ & 0.9769(4) & 11.307(5) & 476.2 & 9.3 &Side lobe                           &$f_{45}$ & 1.8690(9) & 21.632(10) & 218.0 & 6.3 &Side lobe\\
$f_{15}$ &39.7939(3) &460.577(4) & 520.0 &21.8 &$44f_{\rm orb}$                     &$f_{46}$ & 3.6108(9) & 41.791(10) & 218.7 & 7.5 &Side lobe\\
$f_{16}$ &37.8243(4) &437.782(4) & 513.3 &17.9 &$f_9-14f_{\rm orb}$                 &$f_{47}$ &54.7902(10) &634.146(11) & 204.9 & 9.5 &$f_3+35f_{\rm orb}$\\
$f_{17}$ & 4.5327(4) & 52.462(5) & 470.5 &15.3 &$5f_{\rm orb}$                      &\boldmath{$f_{48}$} &\bf{47.2247(10)} &\bf{546.582(11)} &\bf{205.9} &\bf{7.2} &\\
$f_{18}$ &26.9967(4) &312.462(5) & 461.7 &11.1 &$f_9-26f_{\rm orb}$                 &$f_{49}$ & 7.2662(10) & 84.099(11) & 201.1 & 7.9 &$8f_{\rm orb}$\\
\boldmath{$f_{19}$} &\bf{29.7751(4)} &\bf{344.619(5)} & \bf{463.5} &\bf{11.9} &     &$f_{50}$ &52.7597(10) &610.644(12) & 193.7 & 7.6 &$f_{42}+25f_{\rm orb}$\\
$f_{20}$ &24.9236(4) &288.467(5) & 462.5 &10.2 &$f_2+f_3$                           &$f_{51}$ &15.3924(10) &178.152(11) & 199.7 & 8.1 &$17f_{\rm orb}$\\
$f_{21}$ &23.2902(4) &269.562(5) & 463.7 &10.8 &$f_1-f_2$                           &$f_{52}$ &54.6365(10) &632.367(12) & 197.2 & 9.0 &$2f_1+f_{17}$\\
$f_{22}$ & 5.4201(4) & 62.732(5) & 443.6 &15.3 &$6f_{\rm orb}$                      &$f_{53}$ & 2.7523(10) & 31.855(11) & 204.4 & 7.4 &Side lobe\\
$f_{23}$ &27.9642(5) &323.660(5) & 411.8 & 9.7 &$f_{19}-2f_{\rm orb}$               &$f_{54}$ &45.4161(11) &525.649(13) & 174.0 & 6.9 &$f_{48}-2f_{\rm orb}$\\
$f_{24}$ &14.4792(5) &167.583(6) & 392.6 &14.8 &$16f_{\rm orb}$                     &$f_{55}$ &52.9408(11) &612.740(13) & 175.1 & 7.1 &$f_3+33f_{\rm orb}$\\
$f_{25}$ &16.2885(5) &188.524(6) & 379.4 &12.4 &$18f_{\rm orb}$                     &$f_{56}$ &47.0574(11) &544.645(13) & 172.6 & 5.8 &$52f_{\rm orb}$\\
$f_{26}$ &36.6668(5) &424.385(6) & 368.4 &11.0 &$f_3+15f_{\rm orb}$                 &$f_{57}$ &52.8261(12) &611.413(14) & 166.2 & 6.5 &$2f_1+3f_{\rm orb}$\\
$f_{27}$ & 0.9392(5) & 10.870(5) & 409.8 & 7.7 &Side lobe                           &$f_{58}$ &42.1619(12) &487.984(14) & 161.3 & 5.8 &$f_3+21f_{\rm orb}$\\
$f_{28}$ & 3.6453(6) & 42.191(7) & 325.0 &11.2 &$4f_{\rm orb}$                      &$f_{59}$ &54.5726(13) &631.627(15) & 152.3 & 6.9 &$f_{42}+27f_{\rm orb}$\\
$f_{29}$ &18.0978(6) &209.466(7) & 320.7 & 8.8 &$20f_{\rm orb}$                     &$f_{60}$ &58.1072(14) &672.538(16) & 144.2 & 5.8 &$f_{48}+12f_{\rm orb}$\\
$f_{30}$ &64.7241(6) &749.122(7) & 315.0 &13.2 &$f_3+46f_{\rm orb}$                 &$f_{61}$ &60.3409(16) &698.390(18) & 128.6 & 6.5 &$f_1+39f_{\rm orb}$\\
$f_{31}$ &37.9861(6) &439.654(7) & 308.7 &11.6 &$42f_{\rm orb}$\\
\hline
\end{tabular}
\end{table*}
Based on the derived photometric solution, we compute the time-resolved theoretical light curves with the binary model, and then obtain the pulsational light variations by subtracting it from the short cadence data. Figure 2 presents light residuals in plots of magnitude versus BJD. The complete residuals is shown in the upper panel and the close-up view of a portion of the residuals  is shown in the lower panel. In order to investigate the pulsation features in details, we carry out a multiple frequency analysis for the light residuals with the Period04 program (Lenz \& Breger 2005). No signals are detected in the high frequency region ($f > 100$ day$^{-1}$), we then perform further frequency extraction in the frequency range of  0-100 day$^{-1}$.

At each step of the iteration, we select the frequency with the highest amplitude and perform a multi-period least-square fit to the data using all frequencies already detected. The data are then pre-whitened and the residuals was used for further analysis until arriving at the empirical threshold of the signal-to-noise ratio S/N = 5.4 (Baran et al. 2015). Finally, a total of 61 frequencies with S/N $>$ 5.4 were detected, which are listed in Table 2. The noises are calculated in the range of 2 day$^{-1}$ around each frequency, and uncertainties of frequencies and amplitudes are calculated based on the treatment proposed by Montgomery $\&$ O'Donoghue (1999) and Kallinger et al. (2008). Figure 3 presents Fourier amplitude spectra of the light residuals, the original spectrum shown in the upper panel and the residual spectrum of the 61 frequencies after pre-whitening shown in the lower panel.

We check those extracted frequencies and search for possible orbital harmonics and liner combination frequencies in the form of $f_k = f_i \pm mf_{\rm orb}$ or $f_k = mf_i + nf_j$ (P{\'a}pics et al. 2012; Kurtz et al. 2015), where $m$ and $n$ are integers, $f_i$ and $f_j$ are the parent frequencies, $f_k$ is the  combination frequency, and $f_{\rm orb}$ = 0.9049 day$^{-1}$. A peak is accepted as a combination if the amplitudes of both parent frequencies are larger than that of the presumed combination term, and the difference between the observed frequency and the predicted frequency is smaller than the frequency resolution 1.5/$\Delta$T= 0.047 day$^{-1}$ (Loumos $\&$ Deeming 1978; Lee et al. 2019). As shown in Table 2, we identify 46 such frequencies. Besides, when two frequencies are closer than 1.5/$\Delta$T= 0.047 day$^{-1}$, the lower amplitude frequency is identified as the possible side lobe. Given that the kepler satellite rotates 90 degree every 93 days, the corresponding frequency is $f_{\rm sat}$ = 0.011 day$^{-1}$ (Hass 2010; Van Cleve et al. 2016; Yang et al. 2018). We also discard the peak $f_4$, which  is two times of the frequency $f_{\rm sat}$. Finally, six confident independent frequencies $f_{1}$, $f_{3}$, $f_{9}$, $f_{19}$, $f_{42}$, and $f_{48}$ are retained and marked in bold face in Table 2.

The six independent frequencies range from 23.1127 day$^{-1}$ to 50.5292 day$^{-1}$. Connecting with physical parameters of the components in Table 1, we identify the primary star as a $\delta$ Scuti pulsator with multi-periodic pulsations.
\section{Stellar Models}
\subsection{\rm Input Physics}
The one-dimensional stellar evolution code Modules for Experiments in Stellar Astrophysics (MESA, version 10398, Paxton et al. 2011, 2013, 2015, 2018) is used to generate theoretical models. In particular, the submodule called "pulse$\_$adipls" was adopted to compute evolutionary models of stars and calculate adiabatic frequencies of their corresponding radial oscillations and nonradial oscillations (Christensen-Dalsgaard 2008; Paxton et al. 2011, 2013, 2015, 2018) .

In the calculations, the 2005 update of the OPAL equation of state tables (Rogers \& Nayfonov 2002) are used. The OPAL opacity tables of Iglesias \& Rogers (1996) are used for high temperatures region, and tables of Ferguson et al. (2005) used in the low temperature region. We adopt "simple$\_$photosphere" for the atmosphere boundary condition, and use the solar metal composition AGSS09 (Asplund et al. 2009) as the initial ingredient in metal. In the convective region, we use the classical mixing length theory of B$\ddot{\rm o}$hm-Vitense (1958) with the solar value of $\alpha$ = 1.9 (Paxton et al. 2013) to treat convection. In addition, effects of the stellar rotation, the convective overshooting, and the element diffusion, as well as magnetic fields on the structure and evolution of the star are not included in this work (more details can refer to the Appendix).
\subsection{\rm Single-star evolutionary models}
\begin{figure}[h]
\includegraphics[width=\textwidth, angle = 0]{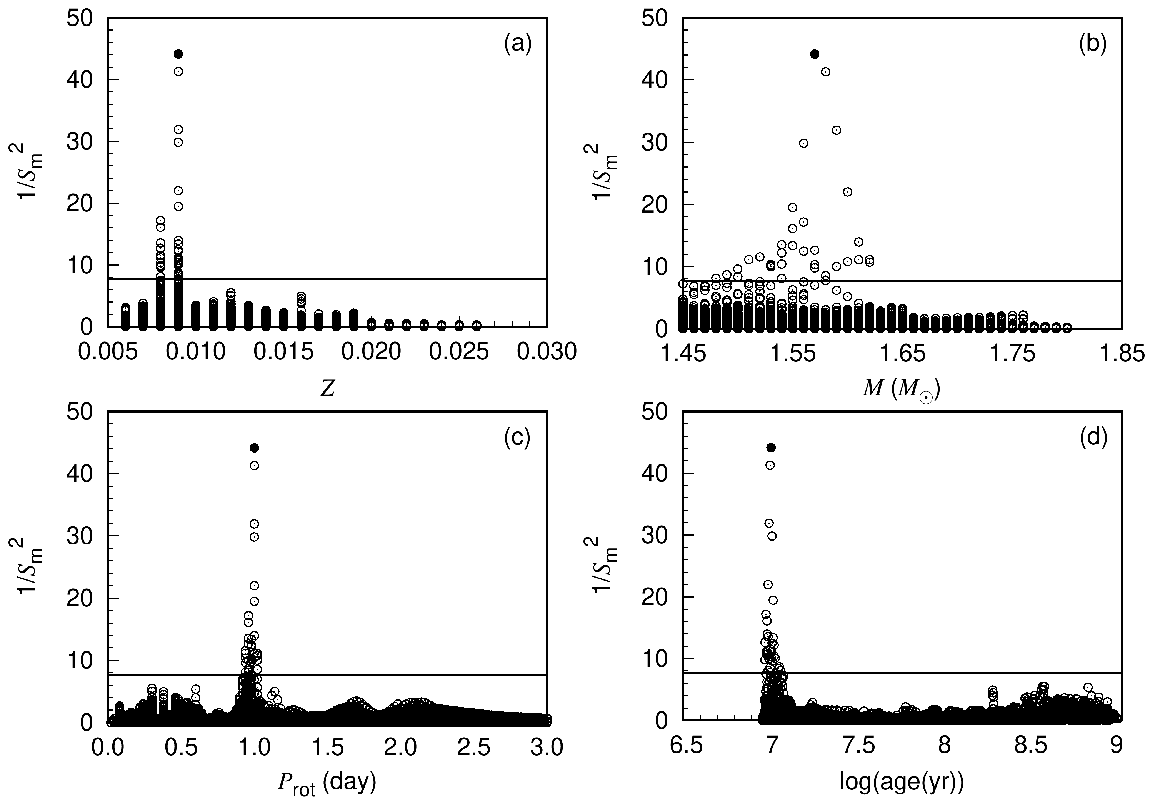}
\caption{\label{Figure4} Visualization of fitting results $S_{\rm m}^2$ versus physical parameters: the metallicity $Z$, the stellar mass $M$, the rotation period $P_{\rm rot}$, and the ages of stars, respectively. The horizontal line in orange marks the position of $S_{\rm m}^2$ = 0.13. The filled circle denotes the best-fitting model.}
\end{figure}
\begin{figure*}[h]
\includegraphics[width=\textwidth, angle = 0]{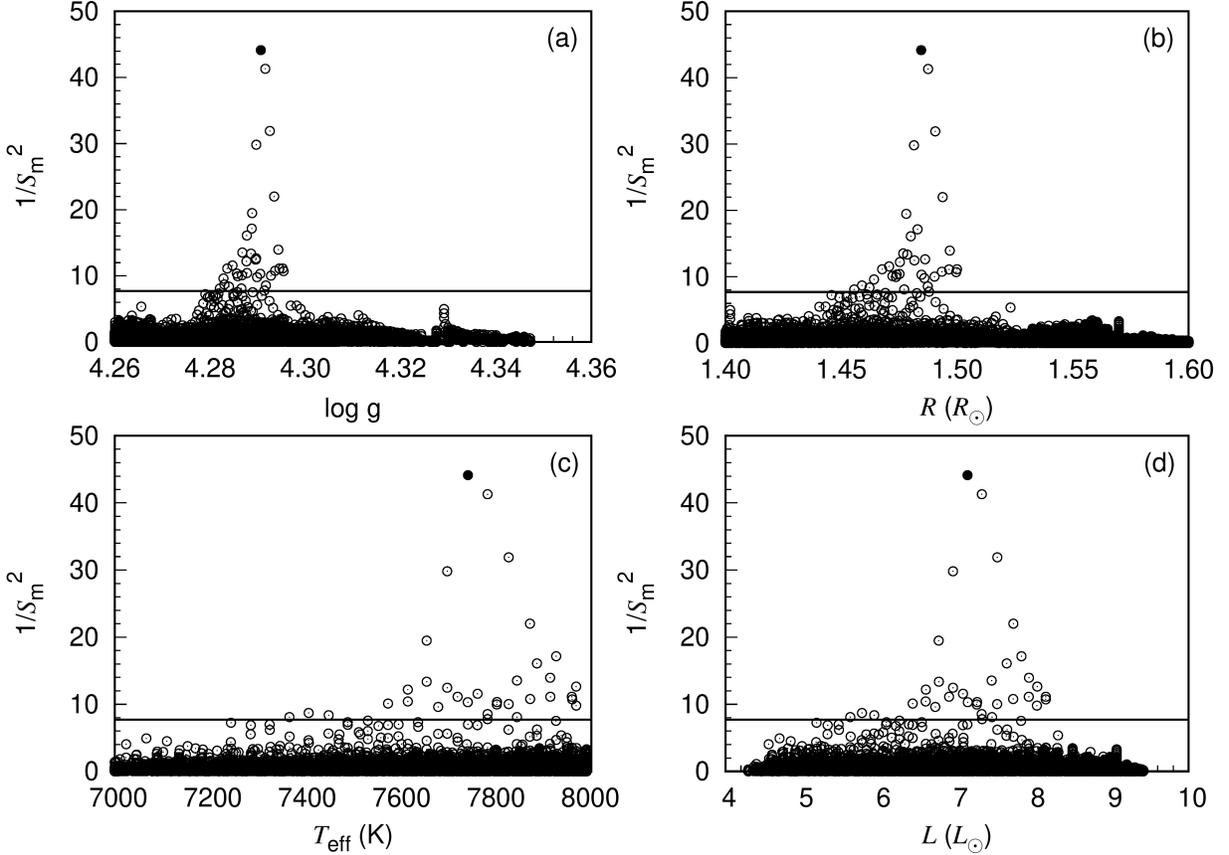}
\caption{\label{Figure5} Visualization of fitting results $S_{\rm m}^2$ versus stellar fundamental parameters: the gravitational acceleration $\log g$, he stellar radius $R$, the effective temperature $T_{\rm eff}$, , and the stellar luminosity $L$, respectively. The horizontal line marks the position of $S_{\rm m}^2$ = 0.13. The filled circle denotes the best-fitting model.}
\end{figure*}
\begin{table*}
\centering
\large
\caption{\label{t3}Candidate models with $S_{\rm m}^{2}$ $\le$ 0.13. $P_{\rm rot}$ is the rotation period. $\tau_0$ is the acoustic radius. $X_{\rm c}$ is the mass fraction of hydrogen in the center of the star. }
\begin{tabular}{ccccccccccccccccc}
\hline\hline
Model &$P_{\rm rot}$ &$Z$  &$M$  &$T_{\rm eff}$ &log$g$   &$R$ &$L$ &$\tau_0$ &$X_{\rm c}$ &Age &$S_{\rm m}^{2}$  \\
&(day)  &     &($M_{\odot}$)  &(K)   &(cgs) &$(R_{\odot})$  &($L_{\odot}$) &(hr)  & &(Myr)  &  \\
\hline
A1  &0.94 &0.008 &1.50 &7685 &4.283 &1.464 &6.718 &1.561 &0.7321 &10.46 & 0.104\\
A2  &0.94 &0.008 &1.51 &7726 &4.284 &1.468 &6.897 &1.563 &0.7321 &10.30 & 0.090\\
A3  &0.94 &0.008 &1.52 &7769 &4.285 &1.471 &7.078 &1.564 &0.7321 &10.16 & 0.087\\
A4  &0.94 &0.008 &1.53 &7810 &4.286 &1.474 &7.261 &1.565 &0.7321 &10.01 & 0.097\\
A5  &0.94 &0.008 &1.54 &7852 &4.287 &1.477 &7.453 &1.566 &0.7321 & 9.87 & 0.124\\
A6  &0.96 &0.008 &1.53 &7810 &4.286 &1.474 &7.256 &1.565 &0.7321 &10.01 & 0.100\\
A7  &0.96 &0.008 &1.54 &7852 &4.287 &1.477 &7.449 &1.566 &0.7321 & 9.87 & 0.074\\
A8  &0.96 &0.008 &1.55 &7895 &4.288 &1.480 &7.644 &1.567 &0.7321 & 9.73 & 0.062\\
A9  &0.96 &0.008 &1.56 &7935 &4.289 &1.483 &7.834 &1.568 &0.7321 & 9.60 & 0.058\\
A10 &0.96 &0.008 &1.57 &7978 &4.290 &1.486 &8.038 &1.569 &0.7320 & 9.46 & 0.079\\
A11 &0.96 &0.009 &1.48 &7370 &4.282 &1.456 &5.617 &1.550 &0.7297 &11.65 & 0.124\\
A12 &0.96 &0.009 &1.49 &7411 &4.283 &1.459 &5.768 &1.551 &0.7297 &11.48 & 0.115\\
A13 &0.96 &0.009 &1.50 &7453 &4.284 &1.462 &5.927 &1.552 &0.7297 &11.31 & 0.119\\
A14 &0.98 &0.008 &1.57 &7978 &4.290 &1.486 &8.034 &1.568 &0.7320 & 9.47 & 0.102\\
A15 &0.98 &0.009 &1.53 &7579 &4.287 &1.472 &6.425 &1.556 &0.7297 &10.84 & 0.099\\
A16 &0.98 &0.009 &1.54 &7621 &4.288 &1.475 &6.596 &1.558 &0.7297 &10.69 & 0.082\\
A17 &0.98 &0.009 &1.55 &7661 &4.289 &1.478 &6.764 &1.559 &0.7297 &10.55 & 0.075\\
A18 &0.98 &0.009 &1.56 &7704 &4.290 &1.482 &6.948 &1.560 &0.7297 &10.41 & 0.080\\
A19 &0.98 &0.009 &1.57 &7748 &4.291 &1.485 &7.138 &1.561 &0.7297 &10.27 & 0.097\\
A20 &0.98 &0.009 &1.58 &7790 &4.291 &1.488 &7.324 &1.562 &0.7297 &10.13 & 0.129\\
A21 &1.00 &0.009 &1.54 &7621 &4.288 &1.475 &6.594 &1.557 &0.7297 &10.69 & 0.096\\
A22 &1.00 &0.009 &1.55 &7661 &4.289 &1.478 &6.761 &1.558 &0.7297 &10.55 & 0.051\\
A23 &1.00 &0.009 &1.56 &7704 &4.290 &1.481 &6.947 &1.560 &0.7297 &10.41 & 0.034\\
A24 &1.00 &0.009 &1.57 &7748 &4.291 &1.484 &7.136 &1.561 &0.7297 &10.27 & 0.023\\
A25 &1.00 &0.009 &1.58 &7790 &4.292 &1.488 &7.321 &1.562 &0.7297 &10.13 & 0.024\\
A26 &1.00 &0.009 &1.59 &7835 &4.293 &1.491 &7.521 &1.563 &0.7297 &10.00 & 0.031\\
A27 &1.00 &0.009 &1.60 &7880 &4.293 &1.494 &7.728 &1.564 &0.7297 & 9.87 & 0.045\\
A28 &1.00 &0.009 &1.61 &7922 &4.294 &1.497 &7.931 &1.565 &0.7296 & 9.74 & 0.072\\
A29 &1.00 &0.009 &1.62 &7968 &4.295 &1.500 &8.150 &1.566 &0.7296 & 9.62 & 0.090\\
A30 &1.02 &0.009 &1.58 &7790 &4.292 &1.487 &7.320 &1.562 &0.7297 &10.14 & 0.117\\
A31 &1.02 &0.009 &1.59 &7835 &4.293 &1.490 &7.518 &1.562 &0.7297 &10.01 & 0.100\\
A32 &1.02 &0.009 &1.60 &7880 &4.294 &1.493 &7.726 &1.564 &0.7297 & 9.88 & 0.093\\
A33 &1.02 &0.009 &1.61 &7923 &4.295 &1.497 &7.929 &1.565 &0.7296 & 9.75 & 0.090\\
A34 &1.02 &0.009 &1.62 &7969 &4.295 &1.500 &8.149 &1.566 &0.7296 & 9.62 & 0.093\\
\hline
\end{tabular}
\end{table*}
\begin{table*}[h]
\centering
\large
\caption{\label{t4}Physical parameters of the primary star for KIC 10736223 obtained by the asteroseismic fit. $P_{\rm rot}$ denotes the rotation period. $\tau_0$ is the acoustic radius. $X_{\rm c}$ is the mass fraction of hydrogen in the center of the star. The age of mass-accreting models in the third column denotes the evolutionary time since the mass accretion ends.}
\begin{tabular}{lllllll}
\hline\hline
Parameters                   &Single-star models                                  &Mass-accreting models \\
\hline
$Z$                                &0.008-0.009 ($0.009 \pm 0.001$)                      &0.008\\
$M$ ($M_{\odot}$)      &1.48-1.62 ($1.57^{+0.05}_{-0.09}$)                      &\nodata\\
$M_1$ ($M_{\odot}$)   &\nodata                                                                  &0.45-0.50\\
$M_2$ ($M_{\odot}$)  &\nodata                                                                  &1.50-1.56\\
$P_{\rm rot}$(day)       &0.94-1.02 ($1.00^{+0.02}_{-0.06}$)                     &0.94-0.96\\
$T_{\rm eff}$ (K)          &7370-7978 ($7748^{+230}_{-378}$)                  &7673-7972\\
$\log g$ (cgs)               &4.282-4.295 ($4.291^{+0.004}_{-0.009}$)          &4.283-4.289\\
$R$ ($R_{\odot}$)        &1.456-1.500 ($1.484^{+0.016}_{-0.028}$)            &1.464-1.484\\
$L$ ($L_{\odot}$)         &5.617-8.150 ($7.136^{+1.014}_{-1.519}$)              &6.678-7.997\\
$\tau_0$ (hr)                   &1.550-1.569 ($1.561^{+0.008}_{-0.011}$)             &1.561-1.569\\
$X_{\rm c}$                   &0.7296-0.7321 (0.7297$^{+0.0024}_{-0.0001}$) &0.7308\\
Age (Myr)                       &9.46-11.65 ($10.27^{+1.38}_{-0.81}$)                   &2.67-3.14\\
\hline
\end{tabular}
\end{table*}
\begin{table*}[h]
\centering
\caption{\label{t5}Theoretical frequencies derived from the optimal single-star evolutionary model (Model A24). $\nu_{\rm mod}$ denotes the model frequency. $\ell$ and $n$ are its spherical harmonic degree and radial order, respectively. $\beta_{\ell,n}$ is the rotational parameters defined as equation (2). }
\label{observed frequencies}
\begin{tabular}{llllll}
\hline\hline
$\nu_{\rm mod}(\ell,n)$ &$\beta_{\ell, n}$ &$\nu_{\rm mod}(\ell,n)$ &$\beta_{\ell, n}$ &$\nu_{\rm mod}(\ell,n)$ &$\beta_{\ell, n}$\\
\hline
248.342(0,0) &  &133.496(1,0)   &0.515   &200.425(2,0) & 0.800 \\
318.170(0,1)   &  &255.103(1,1)   & 0.989   &256.119(2,0) & 0.991 \\
385.157(0,2) &   &332.967(1,2)  & 0.991   &308.923(2,1) & 0.830 \\
455.871(0,3)  &  &412.580(1,3)  & 0.988   &368.927(2,2) & 0.856 \\
532.781(0,4) &  &492.669(1,4)  & 0.984   &444.292(2,3) & 0.923 \\
613.349(0,5) &  &573.554(1,5)   & 0.980   &524.448(2,4) & 0.951 \\
694.314(0,6) &  &654.779(1,6)   & 0.979   &606.248(2,5) & 0.964 \\
774.716(0,7)  &  &735.115(1,7)    & 0.979   &687.592(2,6) & 0.972 \\
\hline
\end{tabular}
\end{table*}
\begin{table*}[h]
\centering
\large
\caption{\label{t6}Comparisons between model frequencies of the optimal single-star evolutionary model (Model A24) and observations. $\nu_{\rm obs}$ is the observed frequency. $\nu_{\rm mod}$ is the model frequency. $|\nu_{\rm obs}-\nu_{\rm mod}|$ denotes the difference between the observed frequency and its model counterpart.}
\begin{tabular}{ccclc}
\hline\hline
ID  &$\nu_{\rm obs}$   &$\nu_{\rm mod}$ &($\ell$, $n$, $m$) & $|\nu_{\rm obs}-\nu_{\rm mod}|$\\
     &($\mu$Hz)         &($\mu$Hz)              &              &($\mu$Hz)\\
\hline
$f_{1}$        &289.867           &289.710       &(2, 1, -2)       &0.157\\
$f_{3}$       &267.508           &267.589       &(2, 0, +1)       &0.081\\
$f_{9}$      &584.829           &584.897       &(1, 5, +1)        &0.068\\
$f_{19}$    &344.619           &344.427       &(1, 2, +1)        &0.192\\
$f_{42}$    &348.891           &349.112        &(2, 2, -2)        &0.221\\
$f_{48}$    &546.582          &546.462      &(2, 4, +2)        &0.120\\
\hline
\end{tabular}
\end{table*}
A grid of single-star evolutionary models are computed with the MESA code. The mass fraction of helium $Y$ in the grid is set to be $Y= 0.249+1.33Z$ (Li et al. 2018), as a function of the mass fraction of heavy elements $Z$. Thus the evolutionary track and interior structure of a star are completely determined by $M$ and $Z$. We consider the stellar mass $M$ between 1.45 $M_{\odot}$ and 2.00 $M_{\odot}$ in a mass interval of 0.01 $M_{\odot}$, and metaillicities $Z$ between 0.003 and 0.030 in a interval of 0.001.

Each star in the grid is computed starting from pre-main sequence stage and ending when the central hydrogen of the star is exhausted ($X_{\rm c} < 1 \times 10^{-5}$). Based on the binary model, we adopt $7000 < T_{\rm eff} < 8000$ K, $4.26 < \log g < 4.40$, and 1.40 $R_{\odot}$ $<R<$ 1.60 $R_{\odot}$ as the observational constraints for potential models. When a star evolves along its evolutionary track into this region, adiabatic frequencies of the radial oscillations ($\ell = 0$) and nonradial oscillations with $\ell$ = 1 and 2 are calculated for the structure model at each evolutionary step. Due to the cancellation effects of the surface geometry, oscillation modes with higher degree are hardly visible, we then do not include oscillation modes with $\ell \ge 3$.

Besides, the component stars in binaries always rotate along with their orbital motion. We introduce the rotation period $P_{\rm rot}$ of the star as another adjustable parameter (Chen et al. 2019), and consider $P_{\rm rot}$ between 0 and 3 days with a step of 0.02 days. Effects of rotation will result in that each nonradial oscillation mode with the spherical harmonic degree $\ell$ splits into $2\ell+1$ different frequencies. The general expression of the first-order effect of rotation on pulsation was derived as
\begin{equation}
\nu_{\ell,n,m}=\nu_{\ell,n}+m\delta\nu_{\ell,n}=\nu_{\ell,n}+\beta_{\ell,n}\frac{m}{P_{\rm rot}}
\end{equation}
(Aerts et al. 2010), where $\delta\nu_{\ell,n}$ is the splitting frequency, and the radial orders $n$, the spherical harmonic degree $\ell$, and the azimuthal number $m$ are three indices characterizing oscillation modes. As shown in equation (1), the effect of the rotation on pulsation is completely determined by the constant $\beta_{\ell,n}$, which is deduced to be
\begin{equation}
\beta_{\ell, n}=\frac{\int_{0}^{R}(\xi_{r}^{2}+L^{2}\xi_{h}^{2}-2\xi_{r}\xi_{h}-\xi_{h}^{2})r^{2}\rho dr}
{\int_{0}^{R}(\xi_{r}^{2}+L^{2}\xi_{h}^{2})r^{2}\rho dr}
\end{equation}
(Aerts et al. 2010). In equation (2), $\xi_r$ represents the radial displacement, $\xi_h$ represents the horizontal displacement, $\rho$ represents the local density of the star, and $L^2= \ell(\ell+1)$. According to equation (2), each dipole mode splits into three different components, corresponding to modes with $m$ = -1, 0, and +1, respectively. Each quadrupole mode splits into five different components, corresponding to modes with $m$ = -2, -1, 0, +1, and +2, respectively.

Then we perform a $S^2$ minimization by comparing freuencies between model and observations according to 
\begin{equation}
S^{2}=\frac{1}{k}\sum(|\nu_{\rm mod,i}-\nu_{\rm obs, i}|^{2}),
\end{equation}
where $k$ is the number of the observed frequencies, $\nu_{\rm mod,i}$ and $\nu_{\rm obs,i}$ are a pair of matched model-observed frequencies.  Due to no preconceived idea of identifications of the six $\delta$ Scuti frequencies $f_1$, $f_3$, $f_9$, $f_{19}$, $f_{42}$, and $f_{48}$, the random fitting algorithm is adopted. The model frequency nearest to the observed frequency is treated as the most probably matched model counterpart. 

Figures 4 and 5 depict changes of fitting results $S_{\rm m}^{2}$ versus various physical parameters. Similar to work of Chen et al. (2016), the solution is found to be limited to a small parameter space for a given evolutionary track, we thus choose a single model with the minimum value of $S^2$ per evolutionary track, and denote the minimun value with $S_{\rm m}^2$. The horizontal lines in the figures mark the position of $S^2_{\rm m}$ = 0.13, which corresponds to the square of $1/ \Delta T$. The circles above the horizontal line correspond to 34 candidate models in Table 3. Model A24 has the minimum value of $S_{\rm m}^2$ = 0.023, we thus consider Model A24 as the best-fitting model and mark it with the filled circles in Figures 4 and 5.

Figures 4(a)-(c) present changes of $S_{\rm m}^{2}$ as a function of the metallicity $Z$, the stellar mass $M$, and the rotation period $P_{\rm rot}$, respectively. In the figures, we find that values of $Z$ and $P_{\rm rot}$ show excellent convergence, i.e., $Z$ = 0.009 $\pm$ 0.001 and $P_{\rm rot}$ = 1.00$^{+0.02}_{-0.06}$ days. However, values of $M$ are found to cover a wide range between 1.48 $M_{\odot}$ and 1.62 $M_{\odot}$.

Figure 4(d) presents changes of $S_{\rm m}^{2}$ as a function of the age of the star. It can be clearly seen in the figure that ages of candidate models converge well to $10.27^{+1.38}_{-0.81}$ Myr. The pulsating primary looks like to be an almost unevolved star on zero-age main sequence, probably indicating that the binary system KIC 10736223 has just undergone a rapid mass-transfer stage.

Figure 5(a)-(d) present changes of $S_{\rm m}^2$ as a function of various fundamental stellar parameters: the gravitational acceleration $\log g$, the stellar radius $R$, the effective temperature $T_{\rm eff}$, and the stellar luminosity $L$, respectively. As illustrated in the figures, values of $\log g$ and $R$ converge well to 4.291$^{+0.004}_{-0.009}$ and 1.484$^{+0.016}_{-0.028}$ $R_{\odot}$, respectively. However, the convergence of $T_{\rm eff}$ and $L$ are relatively worse, i.e., $T_{\rm eff}$ = 7748$^{+230}_{-378}$ K and $L$ = 7.136$^{+1.014}_{-1.519}$ $L_{\odot}$.

Based on the above analyses, stellar parameters of the primary star obtained by asteroseismology are listed in the second column of Table 4. Those parameters are in agreement with those derived from the binary model. Table 5 lists model frequencies of the best-fitting model. Table 6 lists comparisons between model frequencies derived from the best-fitting model and observations. According to the comparisons, $f_{9}$ and $f_{19}$ are identified as two dipole modes, and $f_{1}$, $f_{3}$, $f_{42}$, and $f_{48}$ as four quadrupole modes.
\subsection{\rm Mass-accreting models}
\begin{figure*}[h]
\includegraphics[width=\textwidth,angle=0]{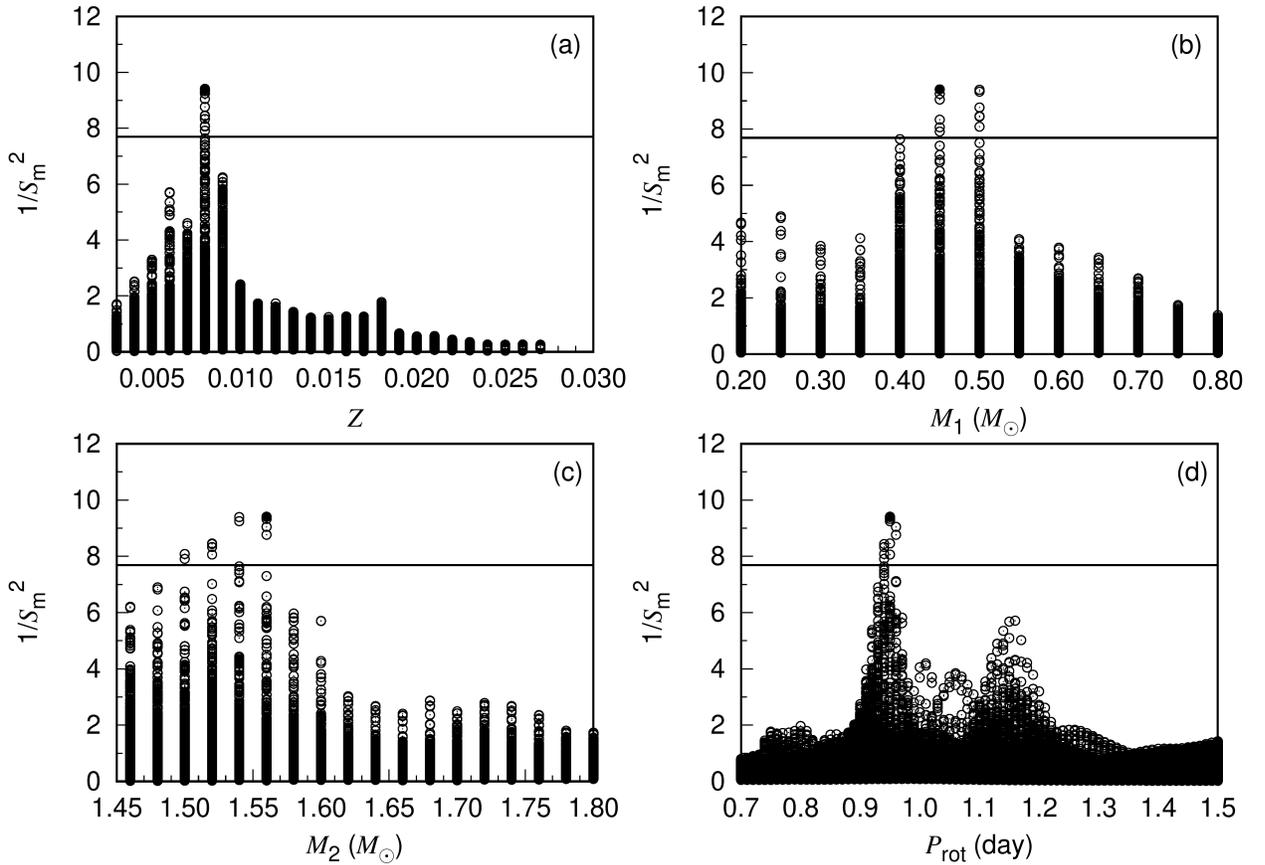}
\caption {\label{Figure6} Visualization of fitting results $S_{\rm m}^2$ versus adjustable parameters of mass-accreting models: the metallicity $Z$, initial stellar mass $M_1$, final stellar mass $M_2$, and the rotation period $P_{\rm rot}$, respectively. The horizontal line marks the position of $S_{\rm m}^2$ = 0.075.}
\end{figure*}
\begin{figure*}[h]
\includegraphics[width=\textwidth, angle = 0]{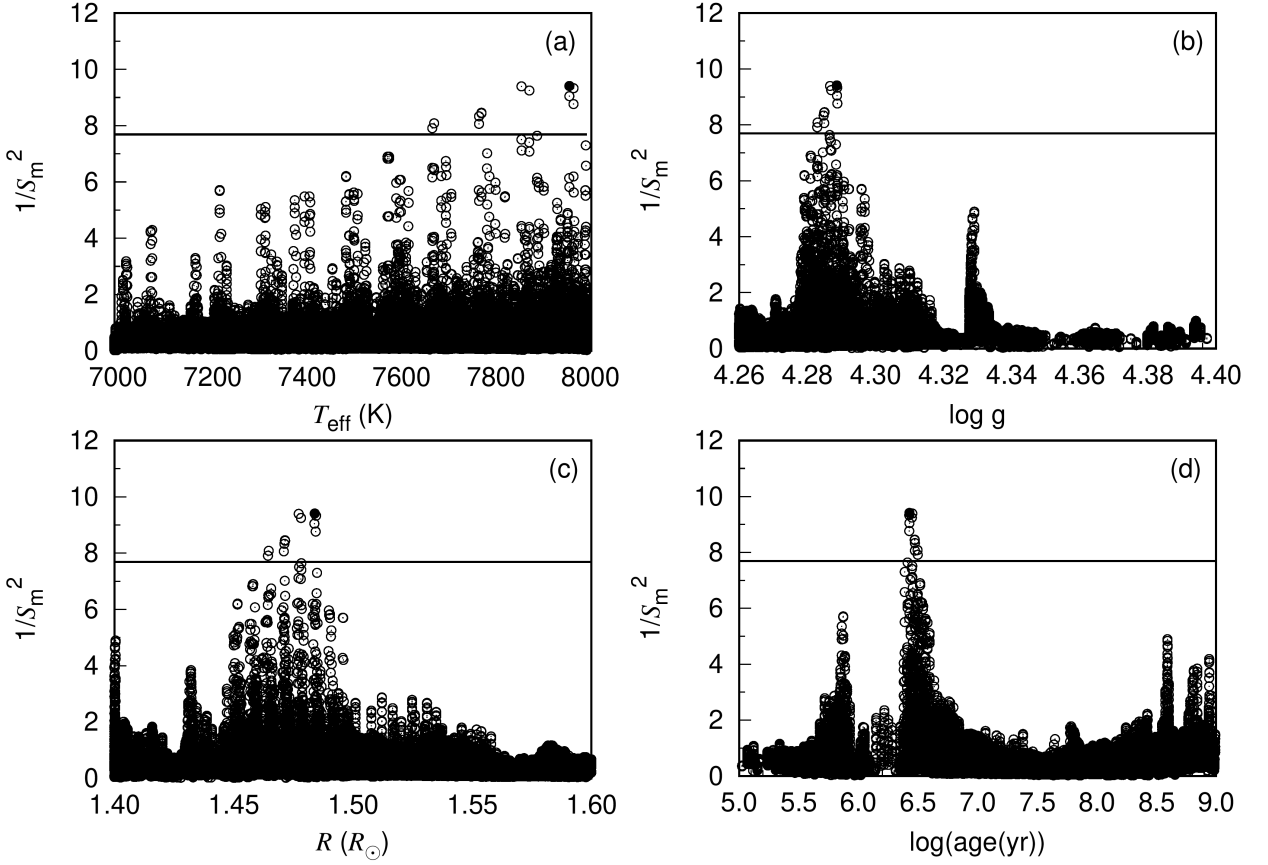}
\caption{\label{Figure7} Visualization of fitting results $S_{\rm m}^2$ versus stellar fundamental parameters of mass-accreting models: the effective temperature $T_{\rm eff}$, the gravitational acceleration $\log g$, he stellar radius $R$, and the age of stars, respectively. The age represents the evolutionary time since the mass accretion ends. The horizontal line marks the position of $S_{\rm m}^2$ = 0.075.}
\end{figure*}
\begin{table*}[h]
\centering
\caption{\label{t7}Candidate mass-accreting models with $S_{\rm m}^{2}$ $\le$ 0.13. $P_{\rm rot}$ is the rotation period. $\tau_0$ is the acoustic radius. $X_{\rm c}$ is the mass fraction of hydrogen in the center of the star. The age represents the evolutionary time since the mass accretion ends.}
\begin{tabular}{ccccccccccccccccc}
\hline\hline
Model &$P_{\rm rot}$ &$Z$  &$M_1$   &$M_2$   &$T_{\rm eff}$ &log$g$    &$R$  &$L$ &$\tau_0$ &$X_{\rm c}$ &Age &$S_{\rm m}^{2}$ \\
&(day)  &     &($M_{\odot}$) &($M_{\odot}$) &(K) &  &$(R_{\odot})$  & ($L_{\odot}$) &(hr)   & &(Myr)  & \\
\hline
B1  &0.94 &0.008 &0.45 &1.50 &7673 &4.283 &1.464 &6.678 &1.561 &0.7308 &3.14 &0.126\\
B2  &0.94 &0.008 &0.45 &1.52 &7772 &4.285 &1.471  &7.095  &1.564 &0.7308 &2.98 &0.120\\
B3  &0.94 &0.008 &0.50 &1.50 &7677 &4.283 &1.465 &6.694 &1.561 &0.7308 &3.13 &0.124\\
B4  &0.94 &0.008 &0.50 &1.52 &7777 &4.285 &1.472  &7.118   &1.564 &0.7308 &2.96 &0.119\\
B5  &0.95 &0.008 &0.45 &1.52 &7772 &4.285 &1.471 &7.093   &1.564 &0.7308 &2.98 &0.124\\
B6  &0.95 &0.008 &0.45 &1.54 &7878 &4.287 &1.478 &7.564  &1.566 &0.7308 &2.81 &0.108\\
B7  &0.95 &0.008 &0.45 &1.56 &7963 &4.289 &1.484 &7.955 &1.569 &0.7308 &2.68 &0.106\\
B8  &0.95 &0.008 &0.50 &1.52 &7777 &4.285 &1.471 &7.117    &1.564 &0.7308 &2.97 &0.118\\
B9  &0.95 &0.008 &0.50 &1.54 &7862 &4.287 &1.477 &7.491  &1.566 &0.7308 &2.83 &0.107\\
B10 &0.95 &0.008 &0.50 &1.56 &7972 &4.289 &1.484 &7.997 &1.569 &0.7308 &2.67 &0.107\\
B11 &0.96 &0.008 &0.45 &1.56 &7963 &4.289 &1.484 &7.955 &1.568 &0.7308 &2.68 &0.111\\
B12 &0.96 &0.008 &0.50 &1.56 &7972 &4.289 &1.484 &7.996 &1.569 &0.7308 &2.67 &0.114\\
\hline
\end{tabular}
\end{table*}
\begin{table*}[h]
\centering
\caption{\label{t8}Comparisons between model frequencies of the optimal mass-accreting model (Model B7) and observations. $\nu_{\rm obs}$ is the observed frequency. $\nu_{\rm mod}$ is the model frequency. $|\nu_{\rm obs}-\nu_{\rm mod}|$ denotes the difference between the observed frequency and its model counterpart.}
\begin{tabular}{ccclc}
\hline\hline
ID  &$\nu_{\rm obs}$   &$\nu_{\rm mod}$ &($\ell$, $n$, $m$) & $|\nu_{\rm obs}-\nu_{\rm mod}|$\\
     &($\mu$Hz)         &($\mu$Hz)              &              &($\mu$Hz)\\
\hline
$f_{1}$    &289.867           &289.912         &(2, 1, -2)           &0.045\\
$f_{3}$    &267.508           &267.098          &(1, 1, 1)            &0.410\\
$f_{9}$    &584.829           &584.475        &(1, 5, +1)        &0.354\\
$f_{19}$    &344.619           &344.826       &(1, 2, +1)        &0.207\\
$f_{42}$    &348.891           &348.763      &(2, 2, -2)          &0.128\\
$f_{48}$    &546.582          &547.114      &(2, 4, +2)        &0.532\\
\hline
\end{tabular}
\end{table*}
KIC 10736223 is a classical Algol binary system formed through mass exchange and mass-ratio reversal. Ideally, binary evolution models with mass transfer should be adopted. However, there are still lots of uncertainties in these models, especially for the mass transfer process. In this work, we modelled the mass transfer process in a simple way. Based on the fitting results of single-star evolutionary models that the binary system probably just past a rapid mass-transfer stage, we first evolved a single star to the position where the rapid mass-transfer begins (the central hydrogen of the donor exhaustion, Case-B binary evolution (Han et al. 2000)), then artifically accreted mass until the mass of the accretor up to the given final values. We considerd thirteen initial stellar masses $M_1$ (the mass accretor) between 0.20 $M_{\odot}$ to 0.80 $M_{\odot}$ with a step of 0.05 $M_{\odot}$. For each initial mass, we evolved the star to the age of 2 Gyr, and then adopt a mass-accreting rate of $10^{-6}$ $M_{\odot}$yr$^{-1}$ at a level of the thermal timescale. We consider the final mass of the accretor between 1.46 $M_{\odot}$ to 1.80 $M_{\odot}$ with a step of 0.02 $M_{\odot}$. For each final mass, we evolve the star until the certral hydrogen of the star is exhausted ($X_{\rm c} < 1 \times 10^{-5}$). Beisdes, we consider $Z$ between 0.003 to 0.030 with a step of 0.001, and $P_{\rm rot}$ between 0.7 days and 1.50 days with a step of 0.01 days.

According to equation (3), we compare frequencies between model and observations. Figures 6 and 7 show changes of fitting results $S_{\rm m}^{2}$ versus various physical parameters of mass-accreting models. The circles above the horizontal line correspond to twelve candidate mass-accreting models in Table 7. The filled circle corresponds to the optimal mass-accreting model (Model B7). In Figures 6(a)-(d), values of $Z$, $M_1$, $M_2$ and $P_{\rm rot}$ converge well to 0.008, 0.45-0.50 $M_{\odot}$, 1.50-1.56 $M_{\odot}$ and 0.94-0.96 days, respectively. In Figures 7(a)-(c), values of $T_{\rm eff}$, $\log g$ and $R$ are also found to cover a small range, i.e., 7673-7972 K, 4.283-4.289 dex and 1.464-1.484 $R_{\odot}$, respectively. As shown in Table 4, the parameters match well with those of the single-star evolutionary models.

Figure 7(d) shows changes of $S_{\rm m}^2$ as a function of the age of the star. Here, the age corresponds to the evolutionary time since the mass accretion ends. It can be clearly seen in the figure that ages of candidate models converge well to $2.68^{+0.46}_{-0.01}$ Myr, also suggesting that the binary system just past the rapid mass transfer stage.

Table 8 lists comparisons between model frequencies of the optimal mass-accreting model and observations. In the table, we notice that identifications of $f_1$, $f_9$, $f_{19}$, $f_{42}$, and $f_{48}$ are the same with those of single-star evolution. For $f_3$, mass-accreting models suggest it to be a dipole mode. 
\section{\rm Summary and Discussions}
We have carried out a detailed analysis for the eclipsing binary KIC 10736223, through binary properties and asteroseismilogy. The results of light curve modeling reveal that the light curve of KIC 10736223 can be almost entirely explained by including a cool spot on the secondary star. Stellar parameters of the two component stars derived from the binary model are $M_1$ = 1.69 $\pm$ 0.09 $M_{\odot}$, $R_1$ = 1.45 $\pm$ 0.03 $R_{\odot}$, $\log g_1$ = 4.34 $\pm$ 0.03, and $M_2$ = 0.34 $\pm$ 0.02 $M_{\odot}$, $R_2$ = 1.35 $\pm$ 0.03 $R_{\odot}$, $\log g_2$ = 3.71 $\pm$ 0.03, respectively. Besides, the simultaneous light-curve and radial-velocity modelling reveals a detached configuration for the binary system with the less-massive secondary nearly filling its Roche lobe.

By subtracting the binary model from the original kepler data, we obtain the light variations due to intrinsic stellar pulsations. Through a multiple frequency analysis for the light residuals, we identify six confident independent frequencies ($f_{1}$, $f_{3}$, $f_{9}$, $f_{19}$, $f_{42}$, and $f_{48}$). These frequencies range from 23.1127 day$^{-1}$ to 50.5292 day$^{-1}$, we then identifiy KIC 10736223 as a member of eclipsing binaries consisting of a $\delta$ Scuti pulsator.

To reproduce the six $\delta$ Scuti frequencies, we compute two grids of theoretical models, including a grid of single-star evolutionary models and a grid of mass-accreting models. Due to no preconceived idea of mode identifications for the observed frequencies, we adopt a random fitting algorithm. Fitting results of mass-accreting models are in good agreement with those of single-star evolutionary models. Fundamental parameters of the primary star are determined to be $M$ = $1.57^{+0.05}_{-0.09}$ $M_{\odot}$, $Z$ = 0.009 $\pm$ 0.001, $R$ = $1.484^{+0.016}_{-0.028}$ $R_{\odot}$, $\log g$ = $4.291^{+0.004}_{-0.009}$, $T_{\rm eff}$ = $7748^{+230}_{-378}$ K, $L$ = $7.136^{+1.014}_{-1.519}$ $L_{\odot}$. According to the theory of stellar oscillations, properties of p modes can be roughly characterized by the acoustic radius $\tau_0$, which is the sound travel time from the surface of the star to the core. Aerts et al. (2010) defined the acoustic radius $\tau_0$ as 
\begin{equation}
\tau_0=\int_0^R\frac{dr}{c_s},
\end{equation}
in which $c_{\rm s}$ is the adiabatic sound speed and $R$ is the stellar radius. As an important asteroseimic parameters, $\tau_0$ is usually used to characterize features of the stellar envelope (e.g., Ballot et al. 2004; Miglio et al. 2010; Chen et al. 2016). Comparing the two kinds of models, we find that their acustic radius $\tau_0$ match well each other, i.e., between 1.50-1.69 hours for single-star evolutionary models and 1.61-1.69 hours for mass-accreting models.

The pulsating primary component of KIC 4544587 is found to be an almost unevolved star near the zore-age main sequence. Ages of the primary star converge to $10.27^{+1.38}_{-0.81}$ Myr for single-star evolutionary models and $2.68^{+0.46}_{-0.01}$ Myr for mass-accreting models, respectively. Given that KIC 10736223 is a classical Algol system formed through mass exchange and mass-ratio reversal, the system KIC 10736223 probably has just past a rapid mass-transfer stage.

Besides, comparisons between model frequencies and observations suggest that $f_9$, $f_{19}$ are two dipole modes and $f_1$, $f_{42}$, $f_{48}$ are three quadrupole modes. For $f_3$, there are two possible identifications, i.e., $\ell$ = 2 by single-star evolutionary models and $\ell$ = 1 by mass-accreting models. Moreover, we find that $f_9$ and $f_{19}$ are identifed as two $m = +1$ dipole modes and $f_1$, $f_{42}$, and $f_{48}$ as three $|m| = 2$ quadrupole modes. This feature reveals that the primary star has a high rotational inclination angle according to Gizon $\&$ Solanki (2003), which meets well with the orbital inclination angle of 89.411 $\pm$ 0.006 degree.

Finally, our asteroseismic results show that the rotation of the primary star is very close to the synchronous rotation, which match well with the previous hypothesis. The rotaion period $P_{\rm rot}$ of the primary star is determined to be $1.00^{+0.02}_{-0.06}$ days for single-star evolutionary models and $0.95\pm0.01$ days for mass-accreting models. According to the work of Saio (1981), Dziembowski \& Goode (1992), and Aerts et al. (2010), the first-order effect of rotation on pulsation is in proportion to $1/P_{\rm rot}$, and that of the second-order is in proportion to $1/(P_{\rm rot}^2\nu_{\ell,n})$. Then their ratio can be estimated to be in the order of $1/(P_{\rm rot}\nu_{\ell,n})$, where $\nu_{\ell,n}$ range from 23.1127 day$^{-1}$ to 50.5292 day$^{-1}$. The second-order effect of rotation on pulsation is much less than that of the first-order one, thus the second-order effect of rotation on pulsation is not included in this work.

\acknowledgments
We are sincerely grateful to the anonymous referee for instructive advice and productive suggestions. This work is supported by the B-type Strategic Priority Program No. XDB41000000 funded by the Chinese Academy of Sciences. The authors also acknowledge supports from the National Natural Science Foundation of China through grants 11803082 to CXH, 11973053 and 11833002 to ZXB, 11333006 and 11521303 to LY,  11703081 to CHL, 11803050 to LCQ, and 11833006 to SJ. Xinghao Chen, Hailiang Chen, and Jie Su also acknowledge the supports of the West Light Foundation of The Chinese Academy of Sciences. Additionally, Hailiang Chen appreciates the support of the Youth Innovation Promotion Association of Chinese Academy of Sciences (Grant no. 2018076). And Changqing Luo appreciates the support of Beijing Natural Science Foundation 1184018. Moreover, the authors gratefully acknowledge the computing time granted by the Yunnan Observatories, and provided on the facilities at the Yunnan Observatories Supercomputing Platform. Finally, Xinghao Chen is thankful for fruitful discussions with Jianning Fu, Dengkai Jiang, and Heran Xiong.
\appendix 
\section{Inlist files used in this work (Version 10398)}
\subsection{\rm The inlsit file for single-star evolution }
\begin{verbatim}
! inlist_pulse
&star_job
 astero_just_call_my_extras_check_model = .true.
 show_log_description_at_start = .false.
 create_pre_main_sequence_model = .true.
 change_lnPgas_flag = .true.
 new_lnPgas_flag = .true.
 change_initial_net = .true.
 new_net_name = 'o18_and_ne22.net'
 kappa_file_prefix = 'a09'
 kappa_lowT_prefix = 'lowT_fa05_a09p'
 initial_zfracs = 6
/ ! end of star_job namelist
&controls
 initial_mass = 1.57
 initial_z = 0.009
 initial_y = 0.26097
 
 MLT_option = 'ML1'  
 mixing_length_alpha = 1.90
 calculate_Brunt_N2 =.true.
 use_brunt_gradmuX_form = .true.
 which_atm_option = 'simple_photosphere'   !default

 max_number_backups = 50
 max_number_retries = 100
 max_model_number = 80000
 history_interval = 1
 max_num_profile_models = 80000
 xa_central_lower_limit_species(1) = 'h1'
 xa_central_lower_limit(1) = 1d-5
 
 use_other_mesh_functions= .true.
 mesh_delta_coeff = 0.9
 M_function_weight = 50
 max_center_cell_dq = 1d-10
 max_allowed_nz = 80000
 varcontrol_target = 1d-5 !for main-sequence models (2d-4 for pre-main sequence models) 
 max_years_for_timestep = 1d6 !for main-sequence models (2d3 for pre-main sequence models)
/ ! end of controls namelist
\end{verbatim}

\subsection{\rm The inlsit file for mass accretion from $M_1$ to $M_2$}
\begin{verbatim}
! inlist_pulse
&star_job
 astero_just_call_my_extras_check_model = .true.
 show_log_description_at_start = .false.
 create_pre_main_sequence_model = .true. 
 save_model_when_terminate = .true.
 save_model_filename = 'final_mass.mod'
 change_lnPgas_flag = .true.
 new_lnPgas_flag = .true.
 change_initial_net = .true.
 new_net_name = 'o18_and_ne22.net'
 kappa_file_prefix = 'a09'
 kappa_lowT_prefix = 'lowT_fa05_a09p'
 initial_zfracs = 6
/ ! end of star_job namelist
&controls
 initial_z = 0.008
 initial_y = 0.25964
 initial_mass = 0.45

 mass_change=1d-6 !if star_age > 2d9 years
 accrete_same_as_surface = .true.  !if star_age > 2d9 years
 star_mass_max_limit =  1.56  !final stellar mass

 MLT_option = 'ML1'  
 mixing_length_alpha = 1.90
 calculate_Brunt_N2 =.true.
 use_brunt_gradmuX_form = .true.
 which_atm_option = 'simple_photosphere'   !default
 
 max_number_backups = 50
 max_number_retries = 100
 max_model_number = 80000
 history_interval = 1
 max_num_profile_models = 80000
 
 use_other_mesh_functions= .true.
 mesh_delta_coeff = 0.90
 M_function_weight =50
 max_center_cell_dq = 1d-10
  max_allowed_nz=80000
 varcontrol_target = 2d-5 
 max_years_for_timestep = 5d6  ! 5d4 if star_age > 2d9 years
/ ! end of controls namelist
\end{verbatim}

\subsection{\rm The inlsit file for evolution of the accreted models }
\begin{verbatim}
! inlist_pulse
&star_job
 astero_just_call_my_extras_check_model = .true.
 show_log_description_at_start = .false.
 load_saved_model = .true.
 saved_model_name = 'final_mass.mod'
 change_lnPgas_flag = .true.
 new_lnPgas_flag = .true.
 change_initial_net = .true.
 new_net_name = 'o18_and_ne22.net'
 kappa_file_prefix = 'a09'
 kappa_lowT_prefix = 'lowT_fa05_a09p'
 initial_zfracs = 6
/ ! end of star_job namelist
&controls
 initial_mass = 1.56
 initial_z = 0.008
 initial_y = 0.25964

 MLT_option = 'ML1'  
 mixing_length_alpha = 1.90
 calculate_Brunt_N2 =.true.
 use_brunt_gradmuX_form = .true.
 which_atm_option = 'simple_photosphere'   !default

 max_number_backups = 50
 max_number_retries = 100
 max_model_number = 80000
 history_interval = 1
 max_num_profile_models = 80000
 xa_central_lower_limit_species(1) = 'h1'
 xa_central_lower_limit(1) = 1d-5

 use_other_mesh_functions= .true.
 mesh_delta_coeff = 0.9
 M_function_weight = 50
 max_center_cell_dq = 1d-10
 max_allowed_nz = 80000
 varcontrol_target = 2d-5
 max_years_for_timestep = 1d6 !for main-sequence models (1d3 for pre-main sequence models)
/ ! end of controls namelist
\end{verbatim}

\end{document}